# Root Estimator of Quantum States

Yu. I. Bogdanov

*OAO "Angstrem", Moscow, Russia*
*E-mail: bogdanov@angstrem.ru*

**Abstract**
The root estimator of quantum states based on the expansion of the psi function in terms of system eigenfunctions followed by estimating the expansion coefficients by the maximum likelihood method is considered. In order to provide statistical completeness of the analysis, it is necessary to perform measurements in mutually complementing experiments (according to the Bohr terminology). Estimation of quantum states by the results of coordinate, momentum, and polarization (spin) measurements is considered. The maximum likelihood technique and likelihood equation are generalized in order to analyze quantum mechanical experiments. The Fisher information matrix and covariance matrix are considered for a quantum statistical ensemble. The constraints on the energy are shown to result in high-frequency noise reduction in the reconstructed state vector. Informational aspects of the problem of division a mixture are studied and a special (quasi- Bayesian) algorithm for solving this problem is proposed. It is shown that the requirement for the expansion to be of a root kind can be considered as a quantization condition making it possible to choose systems described by quantum mechanics from all statistical models consistent, on average, with the laws of classical mechanics.

**Introduction.**

In the previous paper ("Quantum Mechanical View of Mathematical Statistics" *quant-ph/0303013*- hereafter, Paper 1), is has been shown that the root density estimator, based on the introducing an object similar to the psi function in quantum mechanics into mathematical statistics, is an effective tool of statistical data analysis. In this paper, we will show that in problems of estimating of quantum states, the root estimators are at least of the same importance as in the problems of classical statistical analysis.

Methodologically, the method considered here essentially differs from other well known methods for estimating quantum states that arise from applying the methods of classical tomography and classical statistics to quantum problems [1-3]. The quantum analogue of the distribution density is the density matrix and the corresponding Wigner distribution function. Therefore, the methods developed so far have been aimed at reconstructing the aforementioned objects in analogy with the methods of classical tomography (this resulted in the term "quantum tomography") [4].

In [5], a quantum tomography technique on the basis of the Radon transformation of the Wigner function was proposed. The estimation of quantum states by the method of least squares was considered in [6]. The maximum likelihood technique was first presented in [7,8]. The version of the maximum likelihood method providing fulfillment of basic conditions imposed of the density matrix (hermicity, nonnegative definiteness, and trace of matrix equal to unity) was given in [9,10]. Characteristic features of all these methods are rapidly increasing calculation complexity with increasing number of parameters to be estimated and ill-posedness of the corresponding algorithms, not allowing one to find correct stable solutions.

The orientation toward reconstructing the density matrix overshadows the problem of estimating more fundamental object of quantum theory, i.e., the state vector (psi function). Formally, the states described by the psi function are particular cases of those described by the density matrix. On the other hand, this is the very special case that corresponds to fundamental laws in Nature and is related to the situation when the state described by a large number of unknown parameters may be stable and estimated up to the maximum possible accuracy.

In Sec. 1, the problem of estimating quantum states is considered in the framework of the mutually complementing measurements (Bohr [11]). The likelihood equation and statistical properties of estimated parameters are studied.

In Sec. 2, the maximum likelihood method is generalized on the case when, along with the condition on the norm of a state, additional constraint on energy is introduced. The latter restriction allows one to suppress noise corresponding to the high-frequency range of the spectrum of a quantum state.

In Sec. 3, the problem of reconstructing the spin state is briefly considered. In the nonrelativistic approximation, the coordinate and spin wave functions are factorized; therefore, the corresponding problems of reconstructing the states can be considered independently in the same approximation.

The mixture separation problem is considered in Sec. 4. It is shown that from the information standpoint it is purposefully to divide initial data into uniform (pure) sets and perform independent estimation of the state vector for each set. A quasi- Bayesian self-consistent algorithm for solving this problem is proposed.

In Sec. 5, it is shown that the root expansion basis following from quantum mechanics is preferable to any other set of orthonormal functions, since the classical mechanical equations are satisfied for averaged quantities according to the Ehrenfest theorems. Thus, the requirement for the probability density to be of the root form plays a role of the quantization condition.

**1. Phase Role. Statistical Analysis of Mutually Complementing Experiments. Statistical Inverse Problem in Quantum Mechanics.**

We have defined in the Paper.1 the psi function as a complex-valued function with the squared absolute value equal to the probability density. From this point of view, any psi function can be determined up to arbitrary phase factor $\exp(iS(x))$. In particular, the psi function can be chosen real-valued. For instance, in estimating the psi function in a histogram basis, the phases of amplitudes (4.2) of the Paper.1, which have been chosen equal to zero, could be arbitrary.

At the same time, from the physical standpoint, the phase of psi function is not redundant. The psi function becomes essentially complex valued function in analysis of mutually complementing (according to Bohr) experiments with micro objects [11].

According to quantum mechanics, experimental study of statistical ensemble in coordinate space is incomplete and has to be completed by study of the same ensemble in another (canonically conjugate, namely, momentum) space. Note that measurements of ensemble parameters in canonically conjugate spaces (e.g., coordinate and momentum spaces) cannot be realized in the same experimental setup.

The uncertainty relation implies that the two-dimensional density in phase space $P(x,p)$ is physically senseless, since the coordinates and momenta of micro objects cannot be measured simultaneously. The coordinate $P(x)$ and momentum $\widetilde{P}(p)$ distributions should be studied separately in mutually complementing experiments and then combined by introducing the psi function. We will consider the so-called sharp measurements resulting in collapse of the wave function (for unsharp measurements, see [12]).

The coordinate-space and momentum-space psi functions are related to each other by the Fourier transform ($\hbar = 1$)

$$\psi(x) = \frac{1}{\sqrt{2\pi}} \int \widetilde{\psi}(p) \exp(ipx) dp , \qquad (1.1)$$

$$\widetilde{\psi}(p) = \frac{1}{\sqrt{2\pi}} \int \psi(x) \exp(-ipx) dx . \qquad (1.2)$$

Consider a problem of estimating an unknown psi function ($\psi(x)$ or $\widetilde{\psi}(p)$) by experimental data observed both in coordinate and momentum spaces. We will refer to this problem as an statistical inverse problem of quantum mechanics (do not confuse it with an inverse problem in the scattering theory). The predictions of quantum mechanics are considered as a direct problem. Thus, we consider quantum mechanics as a stochastic theory, i.e., a theory describing statistical (frequency) properties of experiments with random events. However, quantum mechanics is a special stochastic theory, since one has to perform mutually complementing experiments (space-time description has to be completed by momentum-energy one) to get statistically full description of a population (ensemble). In order for various representations to be mutually consistent, the theory should be expressed in terms of probability amplitude rather than probabilities themselves.

A simplified approach to the inverse statistical problem, which will be exemplified by numerical example, may be as follows. Assume that density estimators $P(x)$ and $\widetilde{P}(p)$ have already been found (e.g., by histogram estimation). It is required to approximate the psi function for a statistical ensemble. Figure 1 shows the comparison between exact densities that could be calculated if the psi function of an ensemble is known (solid line), and histogram estimators obtained in mutually complementing experiments. In each experiment, the sample size is 10000 points. In Fig. 2, the exact psi function is compared to that estimated by samples. The solution was found by iteration procedure of adjusting the phase of psi function in coordinate and momentum representations. In zero-order approximation ($(r=0)$), the phases were assumed to be zero. The momentum-space phase in the $r+1$ approximation was determined by the Fourier transform of the psi function in the $r$ approximation in the coordinate space and vice versa.

The histogram density estimator results in the discretization of distributions, and hence, natural use of the discrete Fourier transform instead of a continuous one.

Examples of mutually complementing experiments that are of importance from the physical point of view are diffraction patterns (for electrons, photons, and any other particles) in the near-field zone (directly downstream of the diffraction aperture) and in the Fraunhofer zone (far from the diffraction aperture). The intensity distribution in the near-field zone corresponds to the coordinate probability distribution; and that in the Fraunhofer zone, the momentum distribution. The psi function estimated by these two distributions describes the wave field (amplitude and phase) directly at the diffraction aperture. The psi function dynamics described by the Schrödinger equation for particles and the Leontovich parabolic equation for light allows one to reconstruct the whole diffraction pattern (in particular, the Fresnel diffraction).

In the case of a particle subject to a given potential (e.g., an atomic electron) and moving in a finite region, the coordinate distribution is the distribution of the electron cloud, and the momentum distribution is detected in a thought experiment where the action of the potential abruptly stops and particles move freely to infinity.

In quantum computing, the measurement of the state of a quantum register corresponds to the measurement in coordinate space; and the measurement of the register state after performing the discrete Fourier transform, the measurement in momentum space. A quantum register involving $n$ qubits can be in $2^n$ states; and correspondingly, the same number of complex parameters is to be estimated. Thus, exponentially large number of measurements of identical registers is required to reconstruct the psi function if prior information about this function is lacking.

From (1.1) and (1.2), we straightforwardly have

$$\int \frac{\partial \psi^*(x)}{\partial x}\frac{\partial \psi(x)}{\partial x}dx = \int p^2 \widetilde{\psi}^*(p)\widetilde{\psi}(p)dp. \tag{1.3}$$

From the standpoint of quantum mechanics, the formula (1.3) implies that the same quantity, namely, the mean square momentum, is defined in two different representations (coordinate and momentum). This quantity has a simple form in the momentum representation, whereas in the coordinate representation, it is rather complex characteristic of distribution shape (irregularity). The corresponding quantity is proportional to the Fisher information on the translation parameter of the distribution center.

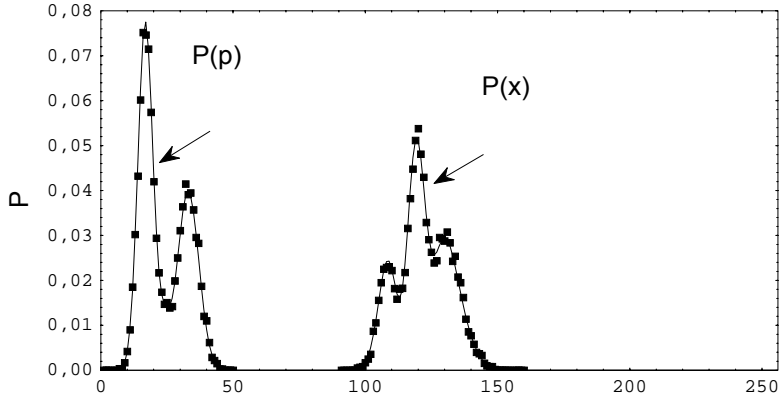

Fig 1. Comparison between exact densities (solid lines) and histogram estimators (dots) in coordinate and momentum spaces.

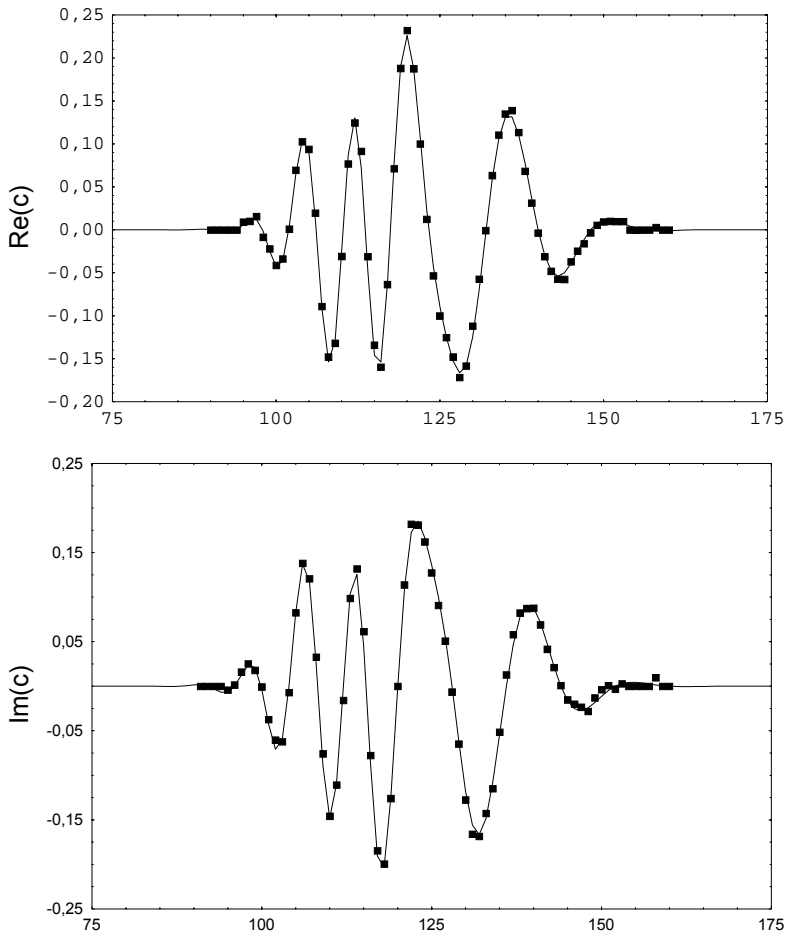

Fig. 2 Comparison between exact psi function (solid line) and that estimated by a sample (dots).

The irregularity cannot be measured in the coordinate space in principle, since it refers to another (momentum) space. In other words, if there is no any information from the canonically conjugate space, the distribution irregularity in the initial space may turn out to be arbitrary high. Singular distributions used in probability theory can serve as density models with an infinite irregularity. From mathematical statistics, it is well-known that arbitrary large sample size does not allow one to determine whether the distribution under consideration is continuous or singular. This causes the ill-posedness of the inverse problem of the probability theory [13].

Thus, from the standpoint of quantum mechanics, the ill-posedness of the classical problem of density estimation by a sample is due to lack of information from the canonically conjugate space. Regularization methods for inverse problem consist in excluding a priory strongly-irregular functions from consideration. This is equivalent to suppression of high momenta in the momentum space.

Let us turn now to more consistent description of the method for estimation of the state vector of a statistical ensemble on the basis of experimental data obtained in mutually complementing experiments. Consider corresponding generalization of the maximum likelihood principle and likelihood equation. To be specific, we will assume that corresponding experiments relate to coordinate and momentum spaces.

We define the likelihood function as

$$L(x,p|c) = \prod_{i=1}^{n} P(x_i|c) \prod_{j=1}^{m} \widetilde{P}(p_j|c). \tag{1.4}$$

Here, $P(x_i|c)$ and $\widetilde{P}(p_j|c)$ are the densities in mutually complementing experiments corresponding to the same state vector $c$. We assume that $n$ measurements were made in the coordinate space; and $m$, in the momentum one.

Then, the log likelihood function has the form (instead of (3.1) of the Paper.1)

$$\ln L = \sum_{i=1}^{n} \ln P(x_i|c) + \sum_{j=1}^{m} \ln \widetilde{P}(p_j|c). \tag{1.5}$$

The maximum likelihood principle together with the normalization condition evidently results in the problem of maximization of the following functional:

$$S = \ln L - \lambda(c_i c_i^* - 1), \tag{1.6}$$

where $\lambda$ is the Lagrange multiplier and

$$\ln L = \sum_{k=1}^{n} \ln(c_i c_j^* \varphi_i(x_k) \varphi_j^*(x_k)) + \sum_{l=1}^{m} \ln(c_i c_j^* \widetilde{\varphi}_i(p_l) \widetilde{\varphi}_j^*(p_l)). \tag{1.7}$$

Here, $\widetilde{\varphi}_i(p)$ is the Fourier transform of the function $\varphi_i(x)$.

The likelihood equation has the form similar to (3.7) of the Paper.1.

$$R_{ij} c_j = \lambda c_i \qquad i,j = 0,1,...,s-1, \tag{1.8}$$

where the $R$ matrix is determined by

$$R_{ij} = \sum_{k=1}^{n} \frac{\varphi_i^*(x_k)\varphi_j(x_k)}{P(x_k)} + \sum_{l=1}^{m} \frac{\widetilde{\varphi}_i^*(p_l)\widetilde{\varphi}_j(p_l)}{\widetilde{P}(p_l)}. \tag{1.9}$$

By full analogy with calculations conducted in Sec.3 of the Paper.1, it can be proved that the most likely state vector always corresponds to the eigenvalue $\lambda = n+m$ of the $R$ matrix (equal to sum of measurements).

The likelihood equation can be easily expressed in the form similar to (3.10) of the Paper.1:

$$\frac{1}{n+m}\left(\sum_{k=1}^{n} \frac{\varphi_i^*(x_k)}{\sum_{j=1}^{s} c_j^* \varphi_j^*(x_k)} + \sum_{l=1}^{m} \frac{\widetilde{\varphi}_i^*(p_l)}{\sum_{j=1}^{s} c_j^* \widetilde{\varphi}_j^*(p_l)}\right) = c_i. \tag{1.10}$$

The Fisher information matrix (prototype) is determined by the total information contained in mutually complementing experiments (compare to (1.5) and (1.6) of the Paper. 1):

$$\widetilde{I}_{ij}(c) = n \cdot \int \frac{\partial \ln P(x,c)}{\partial c_i} \frac{\partial \ln P(x,c)}{\partial c_j^*} P(x,c)dx + m \cdot \int \frac{\partial \ln \widetilde{P}(p,c)}{\partial c_i} \frac{\partial \ln \widetilde{P}(p,c)}{\partial c_j^*} \widetilde{P}(p,c)dp, \quad (1.11)$$

$$\widetilde{I}_{ij} = n \cdot \int \frac{\partial \psi(x,c)}{\partial c_i} \frac{\partial \psi^*(x,c)}{\partial c_j^*} dx + m \cdot \int \frac{\partial \widetilde{\psi}(p,c)}{\partial c_i} \frac{\partial \widetilde{\psi}^*(p,c)}{\partial c_j^*} dp = (n+m)\delta_{ij}. \quad (1.12)$$

Note that the factor of 4 is absent in (1.12) in contrast to the similar formula (1.6) of the Paper.1. This is because of the fact that it is necessary to distinguish $\psi(x)$ and $\psi^*(x)$ as well as $c$ and $c^*$.

Consider the following simple transformation of a state vector that is of vital importance (global gauge transformation). It is reduced to multiplying the initial state vector by arbitrary phase factor:

$$c' = \exp(i\alpha)c, \quad (1.13)$$

where $\alpha$ is arbitrary real number.

One can easily verify that the likelihood function is invariant against the gauge transformation (1.13). This implies that the state vector can be estimated by experimental data up to arbitrary phase factor. In other words, two state vectors that differ only in a phase factor describe the same statistical ensemble. The gauge invariance, of course, also manifests itself in theory, e.g., in the gauge invariance of the Schrödinger equation.

The variation of a state vector that corresponds to infinitesimal gauge transformation is evidently

$$\delta c_j = i\alpha\, c_j \qquad j = 0,1,\ldots,s-1, \quad (1.14)$$

where $\alpha$ is a small real number.

Consider how the gauge invariance has to be taken into account in considering statistical fluctuations of the components of a state vector. The normalization condition ($c_j c_j^* = 1$) yields that the variations of the components of a state vector satisfy the condition

$$c_j \delta c_j^* + (\delta c_j)c_j^* = 0. \quad (1.15)$$

Here, $\delta c_j = \hat{c}_j - c_j$ is the deviation of the state estimator found by the maximum likelihood method from the true state vector characterizing the statistical ensemble.

In view of the gauge invariance, let us divide the variation of a state vector into two terms $\delta c = \delta_1 c + \delta_2 c$. The first term $\delta_1 c = i\alpha\, c$ corresponds to gauge arbitrariness, and the second one $\delta_2 c$ is a real physical fluctuation.

An algorithm of dividing of the variation into gauge and physical terms can be represented as follows. Let $\delta c$ be arbitrary variation meeting the normalization condition. Then, (1.15) yields $(\delta c_j)c_j^* = i\varepsilon$, where $\varepsilon$ is a small real number.

Dividing the variation $\delta c$ into two parts in this way, we have

$$(\delta c_j)c_j^* = (i\alpha c_j + \delta_2 c_j)c_j^* = i\alpha + (\delta_2 c_j)c_j^* = i\varepsilon. \quad (1.16)$$

Choosing the phase of the gauge transformation according to the condition $\alpha = \varepsilon$, we find

$$(\delta_2 c_j)c_j^* = 0. \quad (1.17)$$

Let us show that this gauge transformation provides minimization of the sum of squares of variation absolute values. Let $(\delta c_j)c_j^* = i\varepsilon$. Having performed infinitesimal gauge transformation, we get the new variation

$$\delta c'_j = -i\alpha c_j + \delta c_j. \tag{1.18}$$

Our aim is to minimize the following expression:

$$\delta c'_j \delta c'^*_j = \left(-i\alpha c_j + \delta c_j\right)\left(i\alpha c^*_j + \delta c^*_j\right) = \delta c_j \delta c^*_j - 2\varepsilon\alpha + \alpha^2 \to \min. \tag{1.19}$$

Evidently, the last expression has a minimum at $\alpha = \varepsilon$.

Thus, the gauge transformation providing separation of the physical fluctuation from the variation achieves two aims.

First, the condition (1.15) is divided into two independent conditions:

$$\left(\delta c_j\right)c^*_j = 0 \text{ and } \left(\delta c^*_j\right)c_j = 0 \tag{1.20}.$$

Here, we have dropped the subscript 2 assuming that the state vector variation is a physical fluctuation free of the gauge component).

Second, this transformation results in mean square minimization of possible variations of a state vector.

Let $\delta c$ be a column vector, then the Hermitian conjugate value $\delta c^+$ is a row vector. Statistical properties of the fluctuations are determined by the quadratic form $\delta c^+ \widetilde{I} \delta c = \sum_{i,j=0}^{s-1} \widetilde{I}_{ij} \delta c_j \delta c^*_i$. In order to switch to independent variables, we will explicitly express a zero component in terms of the others. According to (1.20), we have $\delta c_0 = -\dfrac{c^*_j \delta c_j}{c^*_0}$. This leads us to $\delta c_0 \delta c^*_0 = \dfrac{c_i c^*_j}{|c_0|^2}\delta c_j \delta c^*_i$. The quadratic form under consideration can be represented in the form $\sum_{i,j=0}^{s-1} \widetilde{I}_{ij} \delta c_j \delta c^*_i = \sum_{i,j=1}^{s-1} I_{ij} \delta c_j \delta c^*_i$, where the true Fisher information matrix has the form (compare to (1.7) of the Paper.1)

$$I_{ij} = (n+m)\left(\delta_{ij} + \dfrac{c_i c^*_j}{|c_0|^2}\right) \qquad i,j = 1,\ldots,s-1, \tag{1.21}$$

where

$$|c_0| = \sqrt{1 - \left(|c_1|^2 + \ldots + |c_{s-1}|^2\right)}. \tag{1.22}$$

The inversion of the Fisher matrix yields the truncated covariance matrix (without zero component). Having calculated covariations with zero components in an explicit form, we finally find the expression for the total covariance matrix that is similar to (1.19) of the Paper.1:

$$\Sigma_{ij} = \overline{\delta c_i \delta c^*_j} = \dfrac{1}{(n+m)}\left(\delta_{ij} - c_i c^*_j\right) \quad i,j = 0,1,\ldots,s-1. \tag{1.23}$$

The Fisher information matrix and covariance matrix are Hermitian. It is easy to see that the covariance matrix (1.23) satisfies the condition similar to (1.21) of the Paper.1:

$$\Sigma_{ij} c_j = 0. \tag{1.24}$$

By full analogy with the reasoning of Sec.1 of the Paper.1, it is readily seen that the matrix (1.23) is the only (up to a factor) Hermitian tensor of the second order that can be constructed from a state vector satisfying the normalization condition.

The formula (1.23) can be evidently written in the form

$$\Sigma = \frac{1}{(n+m)}(E-\rho), \tag{1.25}$$

where $E$ is the $s \times s$ unit matrix, and $\rho$ is the density matrix.

In the diagonal representation

$$\Sigma = UDU^+, \tag{1.26}$$

where $U$ is the unitary matrix, and $D$ is the diagonal matrix.

The diagonal of the $D$ matrix has the only zero element (the corresponding eigenvector is the state vector). The other diagonal elements are equal to $\frac{1}{n+m}$ (the corresponding eigenvectors and their linear combinations form subspace that is orthogonal complement to a state vector).

The chi-square criterion determining whether the scalar product between the estimated and true vectors $cc^{*(0)}$ is close to unity that is similar to (2.1) of the Paper. 1 is

$$(n+m)\left(1 - \left|cc^{*(0)}\right|^2\right) = \widetilde{\chi}^2_{s-1} = \frac{\chi^2_{2(s-1)}}{2}, \tag{1.27}$$

where $\widetilde{\chi}^2_{s-1}$ is a random variable of the chi-square type related to complex variables of the Gaussian type and equal to half of the ordinary random chi-square of the number of the degrees of freedom doubled.

In terms of the density matrix, the expression (1.27) may be represented in the form

$$\frac{(n+m)}{2} Tr\left(\left(\rho - \rho^{(0)}\right)^2\right) = \widetilde{\chi}^2_{s-1} = \frac{\chi^2_{2(s-1)}}{2} \tag{1.28}$$

Let the density matrix represented by a mixture of two components be

$$\rho = \frac{N_1}{N}\rho^{(1)} + \frac{N_2}{N}\rho^{(2)}, \tag{1.29}$$

where

$N_1 = n_1 + m_1$ and $N_2 = n_2 + m_2$ are the sizes of the first and second samples, respectively, $N = N_1 + N_2$,

$\rho^{(1)}_{ij} = c^{(1)}_i c^{(1)*}_j$,  $\rho^{(2)}_{ij} = c^{(2)}_i c^{(2)*}_j$,

$c^{(1)}_i$ and $c^{(2)}_i$ are the empirical state vectors of the samples.

The chi-square criterion to test the homogeneity of two samples under consideration can be represented in two different forms:

$$\frac{N_1 N_2}{N_1 + N_2}\left(1 - \left|c^{(1)} c^{*(2)}\right|^2\right) = \widetilde{\chi}^2_{s-1} = \frac{\chi^2_{2(s-1)}}{2} \tag{1.30}$$

$$\frac{N_1 N_2}{2(N_1 + N_2)} Tr\left(\left(\rho^{(2)} - \rho^{(1)}\right)^2\right) = \widetilde{\chi}^2_{s-1} = \frac{\chi^2_{2(s-1)}}{2} \tag{1.31}$$

This method is illustrated in Fig. 3. In this figure, the density estimator is compared to the true densities in coordinate and momentum spaces.

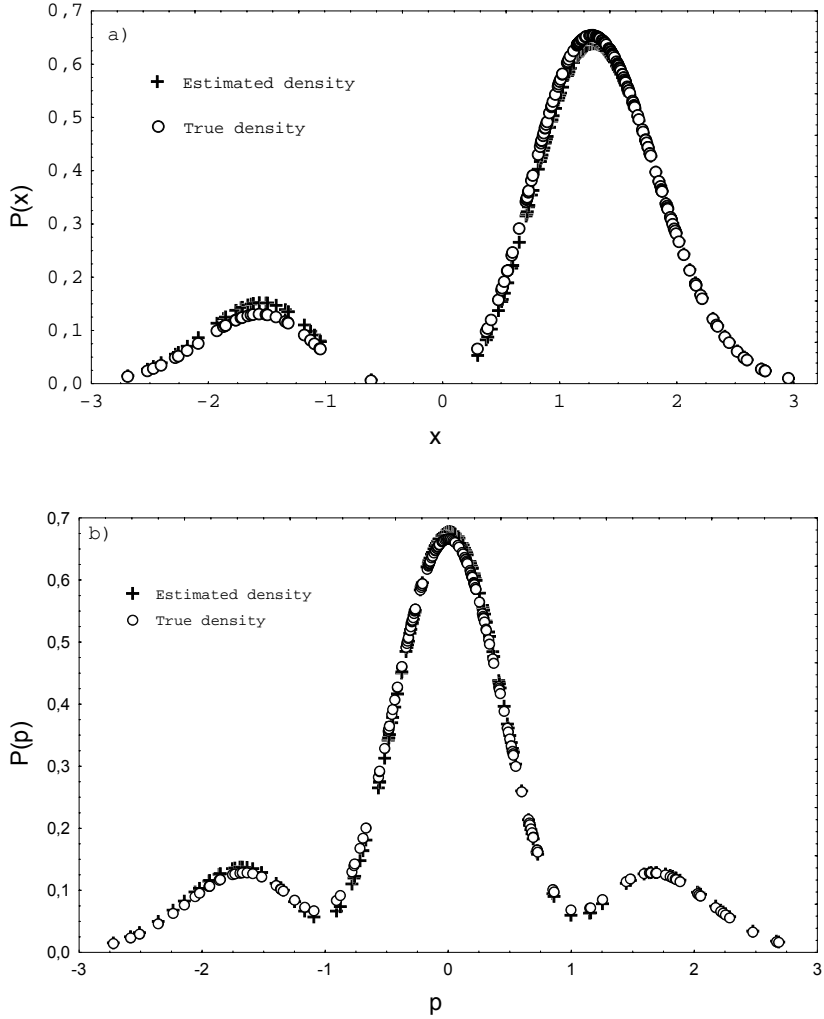

Fig. 3. Comparison between density estimators and true densities in (a) coordinate and (b) momentum spaces

The sample of a size of $n = m = 200$ ($n + m = 400$) was taken from a statistical ensemble of harmonic oscillators with a state vector with three nonzero components ($s = 3$).

The state of quantum register is determined by the psi function

$$\psi = c_i |i\rangle \quad (1.32)$$

The probability amplitudes in the conjugate space corresponding to additional dimensions are

$$\widetilde{c}_i = U_{ij} c_j \quad (1.33)$$

The likelihood function relating to $n + m$ mutually complementing measurements is

$$L = \prod_i \left( c_i c_i^* \right)^{n_i} \prod_j \left( \widetilde{c}_j \widetilde{c}_j^* \right)^{m_j} \quad (1.34)$$

Here, $n_i$ and $m_j$ are the number of measurements made in corresponding states.

In the case under consideration, the likelihood equation similar to (1.10) has the form

$$\frac{1}{n+m} \left[ \frac{n_i}{c_i^*} + \sum_j \frac{m_j U_{ji}^*}{\widetilde{c}_j^*} \right] = c_i \quad (1.35)$$

## 2. Constraint on the Energy

As it has been already noted, the estimation of a state vector is associated with the problem of suppressing high frequency terms in the Fourier series. In this section, we consider the regularization method based on the law of conservation of energy. Consider an ensemble of harmonic oscillators (although formal expressions are written in general case). Taking into account the normalization condition without any constraints on the energy, the terms may appear that make a negligible contribution to the norm but arbitrary high contribution to the energy. In order to suppress these terms, we propose to introduce both constraints on the norm and energy in the maximum likelihood method. The energy is readily estimated by the data of mutually complementing experiments.

It is worth noting that in the case of potentials with a finite number of discrete levels in quantum mechanics [14], the problem of truncating the series does not arise (if solutions bounded at infinity are considered).

We assume that the psi function is expanded in a series in terms of eigenfunctions of the energy operator $\hat{H}$ (Hamiltonian):

$$\psi(x) = \sum_{i=0}^{s-1} c_i \varphi_i(x), \qquad (2.1)$$

where basis functions satisfy the equation

$$\hat{H}\varphi_i(x) = E_i \varphi_i(x). \qquad (2.2)$$

Here, $E_i$ is the energy level corresponding to $i$-th state.

The mean energy corresponding to a statistical ensemble with a wave function $\psi(x)$ is

$$\overline{E} = \int \psi^*(x) \hat{H} \psi(x) dx = \sum_{i=0}^{s-1} E_i c_i^* c_i. \qquad (2.3)$$

In arbitrary basis

$$\overline{E} = \sum_{i=0}^{s-1} H_{ij} c_j c_i^*, \text{ where } H_{ij} = \int \varphi_i^*(x) \hat{H} \varphi_j(x) dx. \qquad (2.4)$$

Consider a problem of finding a maximum likelihood estimator of a state vector in view of a constraint on the energy and norm of the state vector. In energy basis, the problem is reduced to maximization of the following functional:

$$S = \ln L - \lambda_1 (c_i c_i^* - 1) - \lambda_2 (E_i c_i^* c_i - \overline{E}), \qquad (2.5)$$

where $\lambda_1$ and $\lambda_2$ are the Lagrange multipliers and $\ln L$ is given by (1.7).
In this case, the likelihood equation has the form

$$R_{ij} c_j = (\lambda_1 + \lambda_2 E_i) c_i, \qquad (2.6)$$

where the $R$ matrix is determined by (1.9).

In arbitrary basis, the variational functional and the likelihood equation have the forms

$$S = \ln L - \lambda_1 (c_i c_i^* - 1) - \lambda_2 (H_{ij} c_j c_i^* - \overline{E}), \qquad (2.7)$$

$$(R_{ij} - \lambda_2 H_{ij}) c_j = \lambda_1 c_i. \qquad (2.8)$$

Having multiplied the both parts of (2.6) (or (2.8)) by $c_i^*$ and summed over $i$, we obtain the same result representing the relationship between the Lagrange multipliers:

$$(n+m) = \lambda_1 + \lambda_2 \overline{E}. \tag{2.9}$$

The sample mean energy $\overline{E}$ (i.e., the sum of the mean potential energy that can be measured in the coordinate space and the mean kinetic energy measured in the momentum space) was used as the estimator of the mean energy in numerical simulations below.

Now, let us turn to the study of statistical fluctuations of a state vector in the problem under consideration. We restrict our consideration to the energy representation in the case when the expansion basis is formed by stationary energy states (that are assumed to be nondegenerate).

Additional condition (2.3) related to the conservation of energy results in the following relationship between the components

$$\delta \overline{E} = \sum_{j=0}^{s-1} \left( E_j c_j^* \delta c_j + E_j \delta c_j^* c_j \right) = 0. \tag{2.10}$$

It turns out that both parts of the equality can be reduced to zero independently if one assumes that a state vector to be estimated involves a time uncertainty, i.e., may differ from the true one by a small time translation. The possibility of such a translation is related to the time-energy uncertainty relation.

The well-known expansion of the psi function in terms of stationary energy states, in view of time dependence, has the form ($\hbar = 1$)

$$\psi(x) = \sum_j c_j \exp(-iE_j(t-t_0)) \varphi_j(x) =$$
$$= \sum_j c_j \exp(iE_j t_0) \exp(-iE_j t) \varphi_j(x) \tag{2.11}$$

In the case of estimating the state vector up to translation in time, the transformation

$$c_j' = c_j \exp(iE_j t_0) \tag{2.12}$$

related to arbitrariness of zero-time reference $t_0$ may be used to fit the estimated state vector to the true one.

The corresponding infinitesimal time translation leads us to the following variation of a state vector:

$$\delta c_j = it_0 E_j c_j. \tag{2.13}$$

Let $\delta c$ be any variation meeting both the normalization condition and the law of energy conservation. Then, from (1.15) and (2.3) it follows that

$$\sum_j (\delta c_j) c_j^* = i\varepsilon_1, \tag{2.14}$$

$$\sum_j (\delta c_j) E_j c_j^* = i\varepsilon_2, \tag{2.15}$$

where $\varepsilon_1$ and $\varepsilon_2$ are arbitrary small real numbers.

In analogy with Sec.1, we divide the total variation $\delta c$ into unavoidable physical fluctuation $\delta_2 c$ and variations caused by the gauge and time invariances:

$$\delta c_j = i\alpha c_j + it_0 E_j c_j + \delta_2 c_j. \tag{2.16}$$

We will separate out the physical variation $\delta_2 c$, so that it fulfills the conditions (2.14) and (2.15) with a zero right part. It is possible if the transformation parameters $\alpha$ and $t_0$ satisfy the following set of linear equations:

$$\left.\begin{array}{l}\alpha + \overline{E} t_0 = \varepsilon_1 \\ \overline{E}\alpha + \overline{E^2} t_0 = \varepsilon_2 \end{array}\right\}. \qquad (2.17)$$

The determinant of (2.17) is the energy variance

$$\sigma_E^2 = \overline{E^2} - \overline{E}^2. \qquad (2.18)$$

We assume that the energy variance is a positive number. Then, there exists a unique solution of the set (2.17). If the energy dissipation is equal to zero, the state vector has the only nonzero component. In this case, the gauge transformation and time translation are dependent, since they are reduced to a simple phase shift.

In full analogy with the reasoning on the gauge invariance, one can show that in view of both the gauge invariance and time homogeneity, the transformation satisfying (2.17) provides minimization of the total variance of the variations (sum of squares of the components absolute values). Thus, one may infer that physical fluctuations are minimum possible fluctuations compatible with the conservation of norm and energy.

Let us make an additional remark about the time translation introduced above. We will estimate the deviation between the estimated and exact state vectors by the difference between their scalar product and unity. Then, one can state that the estimated state vector is closest to the true one specified in the time instant that differs from the "true" one by a random quantity. In other words, a quantum ensemble can be considered as a time meter, statistical clock, measuring time within the accuracy up to the statistical fluctuation asymptotically tending to zero with increasing number of particles of the system. Time measurement implies the comparison between the readings of real and reference systems. The dynamics of the state vector of a reference system corresponds to an infinite ensemble and is determined by the solution of the Schrödinger equation. The estimated state vector of a real finite system is compared with the "world-line" of an ideal vector in the Hilbert space, and the time instant when the ideal vector is closest to the real one is considered as the reading of the statistical clock. Note that ordinary clock operates in the similar way: their readings correspond to the situation when the scalar product of the real vector of a clock hand and the reference one making one complete revolution per hour reaches maximum value.

Assuming that the total variations are reduced to the physical ones, we assume hereafter that

$$\sum_j (\delta c_j) c_j^* = 0, \qquad (2.19)$$

$$\sum_j (\delta c_j) E_j c_j^* = 0. \qquad (2.20)$$

The relationships found yield (in analogy with Sec.1) the conditions for the covariance matrix $\Sigma_{ij} = \overline{\delta c_i \delta c_j^*}$:

$$\sum_j (\Sigma_{ij} c_j) = 0, \qquad (2.21)$$

$$\sum_j (\Sigma_{ij} E_j c_j) = 0. \qquad (2.22)$$

Consider the unitary matrix $U^+$ with the following two rows (zero and first):

$$\left(U^+\right)_{0j} = c_j^*, \quad (2.23)$$

$$\left(U^+\right)_{1j} = \frac{\left(E_j - \overline{E}\right)c_j^*}{\sigma_E} \quad j = 0,1,\ldots,s-1. \quad (2.24)$$

This matrix determines the transition to principle components of the variation

$$U_{ij}^+ \delta c_j = \delta f_i. \quad (2.25)$$

According to (2.19) and (2.20), we have $\delta f_0 = \delta f_1 = 0$ identically in new variables so that there remain only $s-2$ independent degrees of freedom.

The inverse transformation is

$$U \delta f = \delta c. \quad (2.26)$$

On account of the fact that $\delta f_0 = \delta f_1 = 0$, one may drop two columns (zero and first) in the $U$ matrix turning it into the factor loadings matrix $L$

$$L_{ij} \delta f_j = \delta c_i \quad i = 0,1,\ldots,s-1; \quad j = 2,3,\ldots,s-1. \quad (2.27)$$

The $L$ matrix has $s$ rows and $s-2$ columns. Therefore, it provides the transition from $s-2$ independent variation principle components to $s$ components of the initial variation.

In principal components, the Fisher information matrix and covariance matrix are given by

$$I_{ij}^f = (n+m)\delta_{ij}, \quad (2.28)$$

$$\Sigma_{ij}^f = \overline{\delta f_i \delta f_j^*} = \frac{1}{(n+m)} \delta_{ij}. \quad (2.29)$$

In order to find the covariance matrix for the state vector components, we will take into account the fact that the factor loadings matrix $L$ differs form the unitary matrix $U$ by the absence of two aforementioned columns, and hence,

$$L_{ik} L_{kj}^+ = \delta_{ij} - c_i c_j^* - \frac{\left(E_i - \overline{E}\right)\left(E_j - \overline{E}\right)}{\sigma_E^2} c_i c_j^*, \quad (2.30)$$

$$\Sigma_{ij} = \overline{\delta c_i \delta c_j^*} = L_{ik} L_{jr}^* \overline{\delta f_k \delta f_r^*} = \frac{L_{ik} L_{kj}^+}{n+m}. \quad (2.31)$$

Finally, the covariance matrix in the energy representation takes the form

$$\Sigma_{ij} = \frac{1}{(n+m)}\left(\delta_{ij} - c_i c_j^*\left(1 + \frac{\left(E_i - \overline{E}\right)\left(E_j - \overline{E}\right)}{\sigma_E^2}\right)\right). \quad (2.32)$$

$i, j = 0,1,\ldots,s-1$

It is easily verified that this matrix satisfies the conditions (2.21) and (2.22) resulting from the conservation of norm and energy.

The mean square fluctuation of the psi function is

$$\int \overline{\delta \psi \delta \psi^*} dx = \int \overline{\delta c_i \varphi_i(x) \delta c_j^* \varphi_j^*(x)} dx = \overline{\delta c_i \delta c_i^*} = Tr(\Sigma) = \frac{s-2}{n+m}. \quad (2.33)$$

The estimation of optimal number of harmonics in the Fourier series, similar to that in Sec.6 of the Paper.1, has the form

$$s_{opt} = \sqrt[r+1]{rf(n+m)},\qquad(2.34)$$

where the parameters $r$ and $f$ determine the asymptotics for the sum of squares of residuals:

$$Q(s) = \sum_{i=s}^{\infty}|c_i|^2 = \frac{f}{s^r}.\qquad(2.35)$$

The norm existence implies only that $r > 0$. In the case of statistical ensemble of harmonic oscillators with existing energy, $r > 1$. If the energy variance is defined as well, $r > 2$.

Figure 4 shows how the constraint on energy decreases high-energy noise. The momentum-space density estimator disregarding the constraint on energy (upper plot) is compared to that accounting for the constraint.

Fig. 4 (a) An estimator without the constraint on energy;
       (b) An estimator with the constraint on energy.

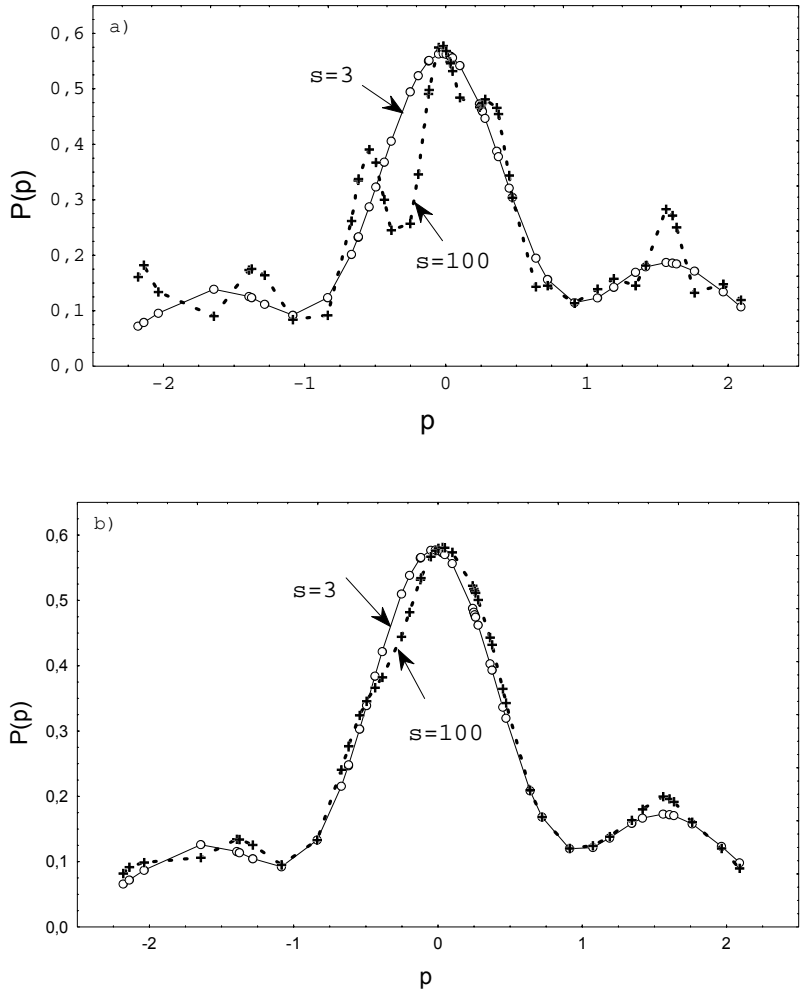

The sample of the size of $n = m = 50$ ($n+m=100$) was taken from the statistical ensemble of harmonic oscillators. The state vector of the ensemble had three nonzero components ($s=3$). Figure 4 shows the calculation results in bases involving $s=3$ and $s=100$ functions, respectively. In the latter case, the number of basis functions coincided with the total sample size. Figure 4 shows that in the case when the constraint on energy was taken into account, the 97 additional noise components influenced the result much weaker than in the case without the constraint.

### 3. Spin State Reconstruction

The aim of this section is to show the application of the approach developed above to reconstruct spin states (by an example of the ensemble of spin-1/2 particles).

We will look for the state vector in the spinor form

$$\psi = c_1 \begin{pmatrix} 1 \\ 0 \end{pmatrix} + c_2 \begin{pmatrix} 0 \\ 1 \end{pmatrix} = \begin{pmatrix} c_1 \\ c_2 \end{pmatrix} \qquad (3.1)$$

The spin projection operator to direction $\vec{n}$ is

$$P(s_n = \pm 1/2) = \frac{1}{2}(1 \pm \vec{\sigma}\vec{n}) \qquad (3.2)$$

We will set $\vec{n}$ by the spherical angles

$$\vec{n} = (n_x, n_y, n_z) = (\sin\theta\cos\varphi, \sin\theta\sin\varphi, \cos\theta) \qquad (3.3)$$

The probabilities for a particle to have positive and negative projections along the $\vec{n}$ direction are

$$P_+(\vec{n}) = \frac{1}{2}\langle\psi|P(s_n = +1/2)|\psi\rangle = \frac{1}{2}\langle\psi|1 + \vec{\sigma}\vec{n}|\psi\rangle \qquad (3.4)$$

$$P_-(\vec{n}) = \frac{1}{2}\langle\psi|P(s_n = -1/2)|\psi\rangle = \frac{1}{2}\langle\psi|1 - \vec{\sigma}\vec{n}|\psi\rangle \qquad (3.5)$$

respectively.

Direct calculation yields the following expressions for the probabilities under consideration:

$$P_+(\vec{n}) = P_+(\theta,\varphi) = \frac{1}{2}\left[(1+\cos\theta)c_1^*c_1 + \sin\theta\, e^{-i\varphi}c_1^*c_2 + \sin\theta\, e^{i\varphi}c_2^*c_1 + (1-\cos\theta)c_2^*c_2\right] \qquad (3.6)$$

$$P_-(\vec{n}) = P_-(\theta,\varphi) = \frac{1}{2}\left[(1-\cos\theta)c_1^*c_1 - \sin\theta\, e^{-i\varphi}c_1^*c_2 - \sin\theta\, e^{i\varphi}c_2^*c_1 + (1+\cos\theta)c_2^*c_2\right] \qquad (3.7)$$

These probabilities evidently satisfy the condition

$$P_+(\theta,\varphi) + P_-(\theta,\varphi) = 1 \qquad (3.8)$$

The likelihood function is given by the following product over all the directions involved in measurements:

$$L = \prod_{\vec{n}} \left(P_+(\vec{n})\right)^{N_+(\vec{n})} \left(P_-(\vec{n})\right)^{N_-(\vec{n})} \qquad (3.9)$$

Here, $N_+(\vec{n})$ and $N_-(\vec{n})$ are the numbers of spins with positive and negative projections along the $\vec{n}$ direction. In order to reconstruct the state vector of a particle, one has to conduct measurements at least in three noncomplanar (linearly independent) directions.

The total number of measurements is

$$N = \sum_{\vec{n}} N(\vec{n}) = \sum_{\vec{n}} \left(N_+(\vec{n}) + N_-(\vec{n})\right) \qquad (3.10)$$

Yu.I. Bogdanov   LANL Report   quant-ph/0303014

In complete analogy to (1.8), we will find the likelihood equation represented by the set of two equations in two unknown complex numbers $c_1$ and $c_2$.

$$\frac{1}{N}\sum_{\vec{n}}\left\{\frac{N_+(\vec{n})\left[(1+\cos\theta)c_1+\sin\theta e^{-i\varphi}c_2\right]}{P_+(\vec{n})\,2}+\frac{N_-(\vec{n})\left[(1-\cos\theta)c_1-\sin\theta e^{-i\varphi}c_2\right]}{P_-(\vec{n})\,2}\right\}=c_1 \quad (3.11)$$

$$\frac{1}{N}\sum_{\vec{n}}\left\{\frac{N_+(\vec{n})\left[\sin\theta e^{i\varphi}c_1+(1-\cos\theta)c_2\right]}{P_+(\vec{n})\,2}+\frac{N_-(\vec{n})\left[-\sin\theta e^{i\varphi}c_1+(1+\cos\theta)c_2\right]}{P_-(\vec{n})\,2}\right\}=c_2 \quad (3.12)$$

This system is nonlinear, since the probabilities $P_+(\vec{n})$ and $P_-(\vec{n})$ depend on the unknown amplitudes $c_1$ and $c_2$.

This system can be easily solved by the method of iterations. The resultant estimated state vector will differ from the true state vector by an asymptotically small random number (the squared absolute value of the scalar product of the true and estimated vectors is close to unity). The corresponding asymptotical formula has the form

$$N\left(1-\left|c^+ c^{(0)}\right|^2\right)=\tilde{\chi}^2_{2j}=\frac{\chi^2_{4j}}{2} \quad (3.13)$$

Here, $\chi^2_{4j}$ is the random number with the chi-square distribution of $4j$ degrees of freedom, and $j$ is the particle spin ($j=1/2$ in our case).

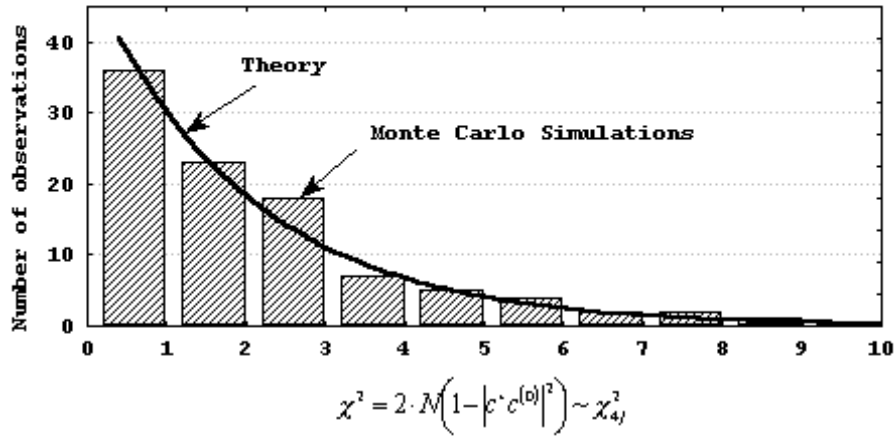

Fig. 5. Spin State Reconstruction
The results of numerical experiments
Chi-Square=2,504    df=5    p=0,776

$\chi^2 = 2\cdot N\left(1-\left|c^+ c^{(0)}\right|^2\right)\sim\chi^2_{4j}$

The validity of the formula (3.13) is illustrated with Fig. 5. In this figure, the results of 100 numerical experiments are presented. In each experiment, the spin $j=1/2$ of a pure ensemble has been measured along 200 random space directions by 50 particles in each direction, i.e., $N=200\cdot 50=10000$. In this case, the left side of (3.13) is the half of a random variable with the chi-square distribution of $4j=2$ degrees of freedom.

Incoherent mixture is described in the framework of the density matrix. Two samples are referred to as mutually incoherent if the squared absolute value of the scalar product of their state vector estimates is statistically significantly different from unity. The quantitative criterion is

$$\frac{N_1 N_2}{N_1 + N_2}\left(1-\left|c^{+(1)}c^{(2)}\right|^2\right) > \widetilde{\chi}^2_{\alpha,2j} = \frac{\chi^2_{\alpha,4j}}{2} \tag{3.14}$$

Here, $N_1$ and $N_2$ are the sample sizes, and $\chi^2_{\alpha,2j}$ is the quantile corresponding to the significance level $\alpha$ for the chi-square distribution of $2j$ degrees of freedom. The significance level $\alpha$ describes the probability of error of the first kind, i.e., the probability to find samples inhomogeneous while they are homogeneous (described by the same psi function).

Let us outline the process of reconstructing states with arbitrary spin. Let $\psi_m^{(j)}$ be the amplitude of the probability to have the projection $m$ along the $z$ axis in the initial coordinate frame (these are the quantities to be estimated by the results of measurements), $m=(j, j-1,\ldots,-j)$. Let $\widetilde{\psi}_m^{(j)}$ be the corresponding quantities in the rotated coordinate frame. The probability to get the value $m$ in measurement along the $z'$ axis is $\left|\widetilde{\psi}_m^{(j)}\right|^2$.

Both rotated and initial amplitudes are related to each other by the unitary transformation

$$\widetilde{\psi}_m^{(j)} = D_{mm'}^{*(j)}\psi_{m'}^{(j)} \tag{3.15}$$

The matrix $D_{mm'}^{(j)}$ is a function of the Euler angles $D_{mm'}^{(j)}(\alpha, \beta, \gamma)$, where the angles $\alpha$ and $\beta$ coincide with the spherical angles of the $z'$ axis with respect to the initial coordinate frame $xyz$, so that $\alpha = \varphi$ and $\beta = \theta$. The angle $\gamma$ corresponds to the additional rotation of the coordinate frame with respect to the $z'$ axis (this rotation is insignificant in measuring the spin projection along the $z'$ axis, and it can be set $\gamma = 0$). The matrix $D_{mm'}^{(j)}$ is described in detail in [14]. Note that our transformation matrix in (3.15) corresponds to the inverse transformation with respect to that considered in [14].

The likelihood equation in this case has the form

$$\frac{1}{N}\sum_{m',\varphi,\theta}\frac{N_{m'}(\varphi,\theta)D_{m'm}^{(j)}(\varphi,\theta)}{\widetilde{\psi}_{m'}^{*(j)}} = \psi_m^{(j)}, \tag{3.16}$$

where $N_{m'}(\varphi,\theta)$ is the number of spins with the projection $m'$ along the $z'$ axis with direction determined by the spherical angles $\varphi$ and $\theta$, and $N$ is the total number of measured spins.

**4. Mixture Division**

There are two different methods for constructing the likelihood function for a mixture resulting in two different ways to estimate the density (density matrix). In the first method (widely used in problems of estimating quantum states [7-10]), the likelihood function for the mixture is constructed regardless a mixed (inhomogeneous) data structure (structureless approach). In this case, for the two-component mixture, we have

$$L_0 = \prod_{i=1}^{n} p\left(x_i \mid c^{(1)}, c^{(2)}, f_1, f_2\right), \tag{4.1}$$

where the mixture density is given by

$$p(x) = f_1 p_1(x) + f_2 p_2(x); \tag{4.2}$$

$p_1(x)$ and $p_2(x)$ are the densities of the mixture components; $f_1$ and $f_2$, their weights.

The normalization condition is

$$f_1 + f_2 = 1 \tag{4.3}$$

Here, it is assumed that all the points of the mixed sample $x_i$ ($i = 1,2,...,n$) are taken from the same distribution $p(x) = f_1 p_1(x) + f_2 p_2(x)$. In other words, the mixture of two inhomogeneous samples (i.e., taken from two different distributions) is treated as a homogeneous sample taken from an averaged distribution.

The second approach, which seems to be more physically adequate, is based on the notion of a mixture as an inhomogeneous population (component approach). This implies that the mixed sample is considered as an inhomogeneous population with $n_1 \approx f_1 n$ points taken from the distribution with the density $p_1(x)$ and the other $n_2 \approx f_2 n$ points, from the distribution with the density $p_2(x)$.

This approach has also formal advantages compared to the first one: it provides higher value of the likelihood function (and hence, higher value of information); besides that, basic theory structures, such as the Fisher information matrix and covariance matrix, take a block form. Thus, the problem is reduced to the division of an inhomogeneous (mixed) population into homogeneous (pure) sub-populations.

In view of the mixed data structure, the likelihood function in the component approach is given by

$$L_1 = \prod_{i=1}^{n_1} p_1\left(x_i \mid c^{(1)}\right) \prod_{i=n_1+1}^{n} p_2\left(x_i \mid c^{(2)}\right). \tag{4.4}$$

Two different cases are possible:
1- population is divided into components a priory from the physical considerations (for instance, $n_1$ values are taken from one source and $n_2$, from another source)
2- dividing the population into components have to be done on the basis of data itself without any information about sources ("blindly")

The first case does not present difficulties, since it is reduced to analyzing homogeneous components in turn. The case when prior information about the sources is lacking requires additional considerations. In order to divide the mixture in this case, we will employ the so-called randomized (mixed) strategy. In this approach, we will consider the observation $x_i$ to be taken from the first distribution with the probability $\dfrac{f_1 p_1(x_i)}{f_1 p_1(x_i) + f_2 p_2(x_i)}$; and from the second one, $\dfrac{f_2 p_2(x_i)}{f_1 p_1(x_i) + f_2 p_2(x_i)}$. Having divided the sample into two population, we will find new estimates of their weights by the formulas $f_1 = \dfrac{n_1}{n}$ and $f_2 = \dfrac{n_2}{n}$, as well as estimates of the state vectors for each population ($c^{(1)}$ and $c^{(2)}$) according to the algorithm presented above. Then, we will find the component densities

$$p_1(x) = \left|c_i^{(1)} \varphi_i(x)\right|^2 \tag{4.5}$$

$$p_2(x) = \left|c_i^{(2)} \varphi_i(x)\right|^2 \tag{4.6}$$

Finally, instead of the initial (prior) estimates of the weights and densities, we will find new (posterior) weights and densities of the mixture components. Applying this (quasi- Bayesian) procedure repeatedly, we will arrive at a certain equilibrium state when the weights and densities of the components become approximately constant (more precisely, prior and posterior estimates of weights and densities become indistinguishable within statistical fluctuations).

A random-number generator should be used for numerical implementation of the algorithm proposed. Each iteration starts with the setting of the random vector of the length $n$ from a homogeneous distribution on the segment [0,1]. If the same random vector is used at each iteration, the distribution of sample points between components will stop varying after some iterations, i.e., each sample point will correspond to a certain mixture component (perhaps, up to insignificant infinite looping, when a periodic exchange of a few points only happens). In this case, each random vector corresponds to a certain random division of mixture into components that allows modeling the fluctuation in a system.

Consider informational aspects of the problem of mixture division into components. The results presented below are based on the following mathematical inequality:

$$f_1 p_1 \ln p_1 + f_2 p_2 \ln p_2 \geq (f_1 p_1 + f_2 p_2) \ln(f_1 p_1 + f_2 p_2), \tag{4.7}$$

that is valid if $f_1 + f_2 = 1$, and $f_1, f_2, p_1$, and $p_2$ are arbitrary nonnegative numbers. We assume also that $0 \ln 0 = 0$.

Generally, the following inequality takes place

$$\sum_{i=1}^{s} (f_i p_i \ln p_i) \geq \sum_{i=1}^{s} (f_i p_i) \ln\left(\sum_{i=1}^{s} (f_i p_i)\right), \tag{4.8}$$

if $\sum_{i=1}^{s} f_i = 1$, and $f_i, p_i$ $(i=1,...,s)$ are arbitrary nonnegative numbers.

The equality sign in (4.8) takes place only in two cases: when either the probabilities are equal to each other ($p_1 = p_2 = ... = p_s$) or one of the weights is equal to unity and the other weights are zero ($f_{i_0} = 1, f_i = 0, i \neq i_0$). In both cases, the mixed state is reduced to the pure one.

The logarithmic likelihood related to a certain observation $x_i$ in the case when we apply the structureless approach and the functional $L_0$ is evidently $\ln(f_1 p_1(x_i) + f_2 p_2(x_i))$. In the component approach when we use the functional $L_1$ and randomized (mixed) strategy, the same observation corresponds to either the logarithmic likelihood $\ln(p_1(x_i))$ with the probability $\dfrac{f_1 p_1(x_i)}{f_1 p_1(x_i) + f_2 p_2(x_i)}$ or the logarithmic likelihood $\ln(p_2(x_i))$ with the probability $\dfrac{f_2 p_2(x_i)}{f_1 p_1(x_i) + f_2 p_2(x_i)}$. In this case, the mean logarithmic likelihood is given by

$$\dfrac{f_1 p_1(x_i)\ln(p_1(x_i)) + f_2 p_2(x_i)\ln(p_2(x_i))}{f_1 p_1(x_i) + f_2 p_2(x_i)}$$

This quantity turns out to be not smaller than the logarithmic likelihood in the first case

$$\dfrac{f_1 p_1(x_i)\ln(p_1(x_i)) + f_2 p_2(x_i)\ln(p_2(x_i))}{f_1 p_1(x_i) + f_2 p_2(x_i)} \geq \ln(f_1 p_1(x_i) + f_2 p_2(x_i)) \qquad (4.9)$$

The validity of the last inequality at arbitrary values of the argument $x_i$ follows from the inequality (4.7). Consider a continuous variable $x$ instead of the discrete one $x_i$ and rewrite the last inequality in the form

$$f_1 p_1(x)\ln(p_1(x)) + f_2 p_2(x)\ln(p_2(x)) \geq$$
$$\geq (f_1 p_1(x) + f_2 p_2(x))\ln(f_1 p_1(x) + f_2 p_2(x)) \qquad (4.10)$$

Integrating with respect to $x$, we find for the Boltzmann $H$ function (representing the entropy with the opposite sign [15])

$$H_{mix} \geq H_0, \qquad (4.11)$$

where

$$H_0 = \int p(x)\ln p(x)\,dx =$$
$$= \int (f_1 p_1(x) + f_2 p_2(x))\ln(f_1 p_1(x) + f_2 p_2(x))\,dx \qquad (4.12)$$

$$H_{mix} = f_1 \int p_1(x)\ln p_1(x)\,dx + f_2 \int p_2(x)\ln p_2(x)\,dx \qquad (4.13)$$

Thus, in the component approach, the Boltzmann $H$ function is higher and the entropy $S = -H$ is lower compared to the structureless description. This means that the representation of data as a mixture of components results in more detailed (and hence, more informative) description than that in the structureless approach. The difference $I_{mix} = H_{mix} - H_0$ can be interpreted as information produced in result of dividing the mixture into components. This information is lost in turning from the component description to structureless.

In general case of arbitrary number of mixture components, the mixture density is

$$p(x) = \sum_{i=1}^{s} f_i p_i(x), \text{ where } \sum_{i=1}^{s} f_i = 1 \qquad (4.14)$$

From inequality (4.8), it follows that the component description generally corresponds to higher (compared to that in the structureless description) value of the Boltzmann $H$ function

$$H_{mix} \geq H_0, \qquad (4.15)$$

where $H_0 = \int p(x)\ln p(x)\,dx \qquad (4.16)$

$$H_{mix} = \sum_{i=1}^{s} f_i H_i, \quad H_i = \int p_i(x) \ln p_i(x) dx \qquad (4.17)$$

The information $I_{mix}$ produced in result of dividing the mixture into components belongs to the range

$$0 \leq I_{mix} \leq S_{sh}, \qquad (4.18)$$

where $S_{sh}$ is the Shanon entropy [16] (the only difference is that we use $e$ (instead of 2) as a base of logarithm)

$$S_{sh} = -\sum_{i=1}^{s}(f_i \ln f_i) \qquad (4.19)$$

The information $I_{mix}$ is equal to zero both in one-component mixture and when the components are indistinguishable (in this case, there is only one nonzero element in the diagonal representation of the density matrix, i.e., it is a pure state). On the contrary, the information $I_{mix}$ reaches its maximum ($I_{mix} = S_{sh}$) when the densities of various components are separated from each other (their ranges of definition do not overlap).

Any density matrix may be represented in the form

$$\rho = LL^+ \qquad (4.20)$$

In the simplest case of second-order density matrix, the matrix $L$ can be represented in the form of expansion

$$L = a_0 E + a_1 \sigma_1 + a_2 \sigma_2 + a_3 \sigma_3, \qquad (4.21)$$

where $E$ is an identity matrix of the second order and $\sigma_1, \sigma_2, \sigma_3$ are the Pauli matrices.

The expansion coefficients are given by

$$a_0 = \frac{1}{2} Tr(L) \qquad a_i = \frac{1}{2} Tr(\sigma_i L) \quad i = 1,2,3 \qquad (4.22)$$

From the normalization condition $Tr(\rho) = 1$, it follows that

$$2(a_0 a_0^* + a_i a_i^*) = 1 \qquad (4.23)$$

As is well known, the generators of the $SU(2)$ group are the spin matrices $\vec{\sigma}/2$. In the general case of $N$-th order density matrix, the expansion similar to (4.21) have to involve the generators of the $SU(N)$ group.

The coordinate distribution density is

$$P(x) = \rho_{ij} \varphi_j^*(x) \varphi_i(x) = L_{il} L_{lj}^+ \varphi_j^*(x) \varphi_i(x) = \psi(x) \psi^+(x). \qquad (4.24)$$

Here, we have introduced the "psi function" in the form of the row matrix

$$\psi_l(x) = \varphi_i(x) L_{il} \qquad (4.25)$$

or

$$\psi(x) = \Phi(x)L, \qquad (4.26)$$

where $\Phi(x) = (\varphi_0(x), \varphi_1(x))$

The expanded form of (4.26) is written as

$$\psi(x) = (\varphi_0(x), \varphi_1(x))(a_0 E + a_i \sigma_i) =$$
$$= b_1(\varphi_0(x), 0) + b_2(\varphi_1(x), 0) + b_3(0, \varphi_0(x)) + b_4(0, \varphi_1(x)) \qquad (4.27)$$

where

$$b_1 = a_0 + a_3, \quad b_2 = a_1 + ia_2, \quad b_3 = a_1 - ia_2, \quad b_4 = a_0 - a_3 \qquad (4.28)$$

Similarly, the momentum distribution density is

$$\widetilde{P}(p) = \widetilde{\psi}(p)\widetilde{\psi}^+(p), \qquad (4.29)$$

where

$$\widetilde{\psi}(p) = (\widetilde{\varphi}_0(p), \widetilde{\varphi}_1(p))(a_0 E + a_i \sigma_i) =$$
$$= b_1(\widetilde{\varphi}_0(p), 0) + b_2(\widetilde{\varphi}_1(p), 0) + b_3(0, \widetilde{\varphi}_0(p)) + b_4(0, \widetilde{\varphi}_1(p)) \qquad (4.30)$$

Here, $\widetilde{\varphi}_0(p)$, $\widetilde{\varphi}_1(p)$ are the Fourier transforms of the functions $\varphi_0(x)$, $\varphi_1(x)$.

The expansions (4.27) and (4.30) show that the reconstruction method that is not based on the prior information about the mixture sources does not allow one to divide the estimates of the parameters $b_1$ and $b_3$, as well as those of $b_2$ и $b_4$. On the contrary, if the estimation method is a priory based on the fact that several observations correspond to the first source, and the other, to the second; we can use the coefficients $b_1$ and $b_2$ to estimate the parameters of the first component, and $b_3$ and $b_4$, for the second one. The problem of estimating the mixture parameters is reduced to reconstructing pure states.

In this paper, major attention is paid to estimating the pure states representing a more fundamental object compared to mixed states. The measurement results together with classical information about either the sources of particles or the environment allow one (in principle) to reduce the study of the density matrix to the study of mixture components representing pure states. It is necessary to keep in mind that dividing mixture into components is not unique. However, the resultant density matrix is the same for any expansion within the statistical fluctuations.

The results that we have found for the second-order density matrices are evidently valid in general case.

In the case when classical information on the sources of particles is partially or totally unavailable, it is purposeful to use quasi- Bayesian algorithm proposed above to divide mixture into components.

The mean trace of the squared deviation between the estimated and true density matrices is (if the division into components is a priory known)

$$\overline{Tr(\rho - \rho^{(0)})^2} = \frac{2(s-1)}{N}, \qquad (4.31)$$

where $N$ is the total sample size and $S$, the dimension of Hilbert space.

If the mixture has to be divided "blindly", the accuracy of estimation somewhat decreases due to fluctuations in the component weights.

## 5. Root estimator and quantum dynamics: statistical root quantization

Let the dynamics of a classical particle be determined by the Hamilton equation

$$\frac{d}{dt}\vec{x} = \frac{1}{m}\vec{p} \tag{5.1}$$

$$\frac{d}{dt}\vec{p} = -\frac{\partial U}{\partial \vec{x}} \tag{5.2}$$

Assume that the mechanical equations are satisfied only for statistically averaged quantities

$$\frac{d}{dt}\overline{\vec{x}} = \frac{1}{m}\overline{\vec{p}} \tag{5.3}$$

$$\frac{d}{dt}\overline{\vec{p}} = -\overline{\frac{\partial U}{\partial \vec{x}}}, \tag{5.4}$$

where the averaged values result from introducing distributions

$$\overline{\vec{x}} = \int P(x)\vec{x}\,dx \tag{5.5}$$

$$\overline{\vec{p}} = \int \widetilde{P}(p)\vec{p}\,dp \tag{5.6}$$

$$\overline{\frac{\partial U}{\partial \vec{x}}} = \int P(x)\frac{\partial U}{\partial \vec{x}}\,dx \tag{5.7}$$

In the expanded form, these averaged equations are evidently written as

$$\frac{d}{dt}\left(\int P(x)\vec{x}\,dx\right) = \frac{1}{m}\left(\int \widetilde{P}(p)\vec{p}\,dp\right) \tag{5.8}$$

$$\frac{d}{dt}\left(\int \widetilde{P}(p)\vec{p}\,dp\right) = -\left(\int P(x)\frac{\partial U}{\partial \vec{x}}\,dx\right) \tag{5.9}$$

Differentiating (5.8) in view of (5.9), we find the averaged Newton's second law of motion

$$\frac{d^2}{dt^2}\left(\int P(x)\vec{x}\,dx\right) = -\frac{1}{m}\left(\int P(x)\frac{\partial U}{\partial \vec{x}}\,dx\right) \tag{5.10}$$

Let us require the density $P(x)$ to admit the root expansion, i.e.,

$$P(x) = |\psi(x)|^2, \qquad (5.11)$$

where

$$\psi(x) = c_j(t)\varphi_j(x) \qquad (5.12)$$

We will search for the time dependence of the expansion coefficients in the form of harmonic dependence

$$c_j(t) = c_{j0}\exp(-i\omega_j t). \qquad (5.13)$$

Then, Eq. (5.10) yields

$$m(\omega_j - \omega_k)^2 c_{j0} c_{k0}^* \langle k|\vec{x}|j\rangle \exp(-i(\omega_j - \omega_k)t) =$$
$$= c_{j0} c_{k0}^* \langle k|\frac{\partial U}{\partial \vec{x}}|j\rangle \exp(-i(\omega_j - \omega_k)t) \qquad (5.14)$$

Here, the summation over recurring indices $j$ and $k$ is meant. The matrix elements in (5.14) are determined by the formulas

$$\langle k|\vec{x}|j\rangle = \int \varphi_k^*(x)\vec{x}\,\varphi_j(x)dx \qquad (5.15)$$

$$\langle k|\frac{\partial U}{\partial \vec{x}}|j\rangle = \int \varphi_k^*(x)\frac{\partial U}{\partial \vec{x}}\,\varphi_j(x)dx \qquad (5.16)$$

In order for the expression (5.14) to be satisfied at any instant of time for arbitrary initial amplitudes, the left and right sides are necessary to be equal for each matrix element. Therefore,

$$m(\omega_j - \omega_k)^2 \langle k|\vec{x}|j\rangle = \langle k|\frac{\partial U}{\partial \vec{x}}|j\rangle \qquad (5.17)$$

This expression is a matrix equation of the Heisenberg quantum dynamics in the energy representation (written in the form similar to that of the Newton's second law of motion). The basis functions and frequencies satisfying (5.17) are the stationary states and frequencies of a quantum system, respectively (in accordance with the equivalence of the Heisenberg and Schrödinger pictures).

Indeed, let us construct the diagonal matrix from the system frequencies $\omega_j$. The matrix under consideration is Hermitian, since the frequencies are real numbers. This matrix is the representation of a Hermitian operator with eigenvalues $\omega_j$, i.e.,

$$H|j\rangle = \hbar\omega_j|j\rangle \qquad (5.18)$$

Let us find an explicit form of this operator. In view of (5.18), the matrix relationship (5.17) can be represented in the form of the operator equation

$$[H[H\vec{x}]] = \frac{\hbar^2}{m}\hat{\partial}U, \qquad (5.19)$$

Yu.I. Bogdanov   LANL Report   quant-ph/0303014

where $\hat{\partial} = \dfrac{\partial}{\partial \vec{x}}$ is the operator of differentiation and $[\ ]$, the commutator.

The Hamiltonian of a system

$$H = -\frac{\hbar^2}{2m}\hat{\partial}^2 + U(x) \qquad (5.20)$$

is the solution of operator equation (5.19).

Thus, if the root density estimator is required to satisfy the averaged classical equations of motion, the basis functions and frequencies of the root expansion cannot be arbitrary, but have to be eigenfunctions and eigenvalues of the system Hamiltonian, respectively.

The relationships providing that the averaged equations of classical mechanics are satisfied for quantum systems are referred to as the Ehrenfest equations [in our case, Eqs. (5.8) and (5.9)]. These equations are insufficient to describe quantum dynamics. As it has been shown above, an additional condition allowing one to transform a classical system into the quantum one (i.e., quantization condition) is actually the requirement for the density to be of the root form.

Thus, if we wish to turn from the rigidly deterministic (Newtonian) description of a dynamical system to the statistical one, it is natural to use the root expansion of the density distribution to be found, since only in this case a stable statistical model can be found. On the other hand, the choice of the root expansion basis determined by the eigenfunctions of the energy operator (Hamiltonian) is not simply natural, but the only possible way consistent with the dynamical laws.

**Conclusions**

Let us state a short summary.

The root state estimator may be applied to analyze the results of experiments with micro objects as a natural instrument to solve the inverse problem of quantum mechanics: estimation of psi function by the results of mutually complementing (according to Bohr) experiments. Generalization of the maximum likelihood principle to the case of statistical analysis of mutually complementing experiments is proposed.

The Fisher information matrix and covariance matrix are considered for a quantum statistical ensemble. It is shown that the constraints on the norm and energy are related to the gauge and time translation invariances. The constraint on the energy is shown to result in the suppression of high-frequency noise in a state vector approximated.

It is shown that the spin wave function can be estimated by the method similar to that used to estimate the coordinate psi function.

It is purposeful to solve the problem of reconstructing the mixed state described by the density matrix by dividing the initial data into homogeneous components. In the case when the prior information is lacking, one should use the self-consistent quasi- Bayesian algorithm proposed in this paper.

It is shown that the requirement for the density to be of the root form is the quantization condition. Actually, one may say about the root principle in statistical description of dynamic systems. According to this principle, one has to perform the root expansion of the distribution density in order to provide the stability of statistical description. On the other hand, the root expansion is consistent with the averaged laws of classical mechanics when the eigenfunctions of the energy operator (Hamiltonian) are used as basis functions. Figuratively speaking, there is no a regular statistical method besides the root one, and there is no regular statistical mechanics besides the quantum one.